\documentclass[pre,aps,twocolumn,showpacs,floatfix]{revtex4}
\usepackage{epsfig}
\usepackage{graphics,graphicx,color,subfigure}
\usepackage{amssymb}
\usepackage{amsmath}
\usepackage{amsfonts}
\usepackage[latin1]{inputenc}
\usepackage{comment}
\usepackage{subfigure}

\newcommand{\lrangle}[1]{\langle{#1}\rangle}
\renewcommand{\vec}{\mathbf}

\newcommand{\vinf}{v_\infty}

\begin{document}
\title{On the origins of scaling corrections in ballistic growth models}
\author{S. G. Alves}
\email{sidiney@ufv.br}
\author{T. J. Oliveira}
\email{tiago@ufv.br}
\author{S. C. Ferreira}
\email{silviojr@ufv.br}
\affiliation{Departamento de F\'isica, Universidade Federal de Vi\c cosa,
36570-000, Vi\c cosa, MG, Brazil}

\date{\today}

\begin{abstract}
We study the ballistic deposition and the grain deposition models on 
two-dimensional substrates. Using the Kardar-Parisi-Zhang (KPZ) ansatz for 
height fluctuations, we show that the main contribution to the intrinsic width, 
which causes strong corrections to the  scaling, comes from the fluctuations in 
the height increments  along deposition events. Accounting for this correction 
in the scaling analysis, we obtained scaling exponents in excellent agreement 
with the KPZ class. We also propose a method to suppress these corrections, which 
consists in divide the surface in bins of size $\varepsilon$ and use only the 
maximal height inside each bin to do the statistics. Again, scaling exponents in 
remarkable agreement with the KPZ class were found. The binning method allowed 
the accurate determination of the height distributions of the ballistic models 
in both growth and steady state regimes, providing the universal underlying 
fluctuations foreseen for KPZ class in 2+1 dimensions. Our results provide 
complete and conclusive evidences that the ballistic model belongs to the KPZ 
universality class in $2+1$ dimensions. Potential applications of the methods 
developed here, in both numerics and experiments, are discussed.
\end{abstract}
\pacs{68.43.Hn, 68.35.Fx, 81.15.Aa, 05.40.-a}

\maketitle

\section{Introduction}

Non-equilibrium dynamics of growing interfaces has attracted much interest in 
several scientific branches such as Physics, Chemistry, Biology and Engineering 
\cite{barabasi,meakin}. A simple and widespread approach to the modeling of 
evolving  surfaces  considers particles in a random flux that irreversibly aggregate to 
the substrate following a given rule. Considering ballistic trajectories for 
particles that aggregate at a first contact with the deposit we have the 
celebrated ballistic deposition (BD) model~\cite{vold}, formerly proposed to 
simulate rock sedimentation, with applications to the  modeling of thin film 
growth at low temperatures~\cite{meakin} and to describe colloidal particle 
deposition at the edges of evaporating drops~\cite{yunker}. A central 
characteristic of ballistic growth models is the lateral growth
that produces a velocity excess. Other models 
exhibiting this property include the Eden model~\cite{eden}, a 
paradigm in the study of curved surfaces, and models where large grains are 
randomly deposited~\cite{tiago1,masoudi}.

The velocity excess is a hallmark of the Kardar-Parisi-Zhang (KPZ) 
universality class~\cite{KPZ}. Therefore, in the hydrodynamic limit, one expects 
that the growth dynamics of ballistic models is described by the KPZ equation
\cite{KPZ}
\begin{equation}
 \frac{\partial h (\vec{x},t)}{\partial t} = \nu \nabla^{2} h + 
 \frac{\lambda}{2} (\nabla h)^{2} + \xi(\vec{x},t),
\label{eqKPZ}
\end{equation}
where terms in the right side accounts, respectively, for the surface tension, 
 local growth in the normal direction and a delta-correlated noise, with 
$\lrangle{\xi(\vec{x},t)}=0$ and 
$\lrangle{\xi(\vec{x},t)\xi(\vec{x}',t')}=2D\delta(t-t')\delta^d(\vec{x}-\vec{x}
')$, associated to the randomness of the deposition process. In $d=1+1$ 
dimensions, the surface height in KPZ systems asymptotically evolves according 
to the ansatz~\cite{krug92,johansson,PraSpo1,PraSpo2}
\begin{equation} 
h \simeq v_{\infty} t + s_{\lambda} (\Gamma t)^{\beta} \chi, 
\label{eqansatz} 
\end{equation} where $v_{\infty}$ is the asymptotic growth velocity, 
$s_{\lambda}$ is the signal of $\lambda$ in the KPZ equation [Eq.~(\ref{eqKPZ})], 
$\Gamma$ is a non-universal constant associated to the amplitude of  the 
interface fluctuations, $\beta$ is the growth exponent, and $\chi$ is a 
stochastic quantity given by Tracy-Widom~\cite{TW1} distributions. This 
conjecture was confirmed in distinct KPZ 
systems~\cite{TakeSano,TakeuchiSP,Alves11,Oliveira12,Alves13,Takeuchi12} besides 
exact solutions of KPZ equation~\cite{SasaSpo1,Amir,CalaDoussal,Imamura}. Recent 
numerical simulations have shown that the KPZ ansatz can be generalized to 
2+1~\cite{Oliveira13R,healyPRL,healyPRE} and higher \cite{alves14} dimensions, 
but the exact forms of the asymptotic distributions of $\chi$ are yet not known.

Although the KPZ equation was initially proposed to explain ballistic deposition 
models, numerical simulations commonly fail to provide a reliable 
connection between them and the KPZ class, mainly in higher dimensions. For 
example, the interface width $W \equiv \sqrt{\lrangle{h^2}_c}$ (here 
$\lrangle{X^n}_c$ represents the $n$th cumulant of $X$) scaling with time $t$ in 
the growth regime ($W\sim t^\beta$ for $t\ll L^z$, where $z=\alpha/\beta$ is the 
dynamic exponent~\cite{barabasi}) and with the system size $L$ in the steady 
state ($W\sim L^\alpha$ for $t\gg L^z$) leads to growth ($\beta$) and roughness 
($\alpha$) exponents smaller than the KPZ values \cite{raissa,fabioBD,vvdensky}. 
In particular, for the BD model in $d=1+1$, exponents in agreement with the KPZ 
ones were obtained through appropriated extrapolations of effective exponents 
\cite{fabioBD} and, more recently, from extremely large-scale simulations 
accessing the regimes where corrections become negligible~\cite{vvdensky}. 
Moreover, recent studies of height distributions have given additional proofs of 
the KPZ universality of Eden and BD models in $d=1+1$ 
\cite{Alves11,Oliveira12,Alves13,Takeuchi12}. For Eden models, scaling exponents 
and height distributions consistent with KPZ class were also found in $d=2+1$ 
\cite{Eden3d, Oliveira13R}. However, for the BD model and also for a grain 
deposition (GD) model~\cite{tiago1} in $d=2+1$ dimensions, strong corrections to 
the scaling were found ~\cite{fabioBD,tiago1,tiago2} and  
evidences of the KPZ class was limited to the collapse of interface width 
distributions~\cite{tiago2} in the steady state.

A correction in the squared interface width $W^2$ for the Eden model~\cite{eden} 
was proposed long ago as a constant additive term in the Family-Vicsek~\cite{FV} 
ansatz, so 
that \cite{wolf}
\begin{equation}
 W^{2} \simeq L^{2\alpha} f \left( \frac{t}{L^{z}} \right) + w_i^2 ,
\label{eqFVg}
\end{equation}
where the first term at the right side accounts for long wavelength 
fluctuations  and $w_i$ is called 
intrinsic width. Corrections consistent with an intrinsic width have been 
observed in many ballistic models~\cite{tiago2,evans,chavez,moro}. 
The intrinsic width was initially attributed to large steps at surface 
~\cite{wolf}, but it was shown that large 
local height gradients is not a sufficient condition for intrinsic
width since other KPZ models presenting local height differences 
comparable to those of Eden and ballistic deposition do not present a relevant
intrinsic width~\cite{tiago2}.

Finite-time corrections observed in several KPZ systems lead to the modified ansatz~\cite{TakeSano,TakeuchiSP,SasaSpo1,Frings, Alves11,Oliveira12,Alves13,Takeuchi12,Oliveira13R}
\begin{equation}
h \simeq v_{\infty} t + s_{\lambda} (\Gamma t)^{\beta} \chi+\eta+\cdots, 
\label{eqansatzcorr} 
\end{equation}
where $\eta$ is, in principle, a model-dependent stochastic 
quantity responsible by a shift in the mean of the scaled variable 
\begin{equation}
 q=\frac{h-\vinf t}{(\Gamma t)^\beta}
 \label{eq:q}
\end{equation}
in relation to the $\chi$ distributions, which vanishes as $t^{-\beta}$. Corrections in 
higher order cumulants of $q$ and, consequently, of $h$  were also
observed but without universal schema~\cite{Oliveira12,Alves13,TakeSano,TakeuchiSP}.

In the present work,  we perform a detailed study of BD and GD models on 
two-dimensional substrates and show that the intrinsic width $w_i$ can be suited in 
terms of the finite-time corrections of the KPZ ansatz, Eq.~(\ref{eqansatzcorr}), and 
the leading contribution to $w_i$ is due to a stochastic component of the local columnar 
growth intrinsic to ballistic growth. More precisely, we show that $w_i^2$ 
is very close to  the variance of the local height increments during 
the deposition process.  Including this variance in the scaling analysis,  exponents 
in striking agreement with the KPZ ones are found. Since large variances in 
height increments are due to narrow-deep valleys in the surface, we also propose 
a method where the surface is constructed considering only the maximal heights 
inside bins of size $\varepsilon$ and show that the intrinsic width is strongly 
reduced, leading to scaling exponents and height distributions in excellent 
agreement with the KPZ class. Our results providing a thorough confirmation of 
the KPZ universality of the ballistic deposition in $d=2+1$ demystify a longstanding 
question which has been chased for decades. Applications of our methods to 
other important ballistic systems are discussed.

The sequence of this paper is organized as follows. The investigated models and the 
method to define the surface are presented in Sec.~\ref{SecModels}. The 
determination of the non-universal parameters in the KPZ ansatz given by 
Eq.~(\ref{eqansatz}) is done in section~\ref{SecPara}. The analysis of the 
scaling corrections and their consequences to the scaling exponents of ballistic 
growth models are presented in Sec.~\ref{SecIntrinsic}. Universality of the 
underlying fluctuations in height distributions is analyzed in 
section~\ref{SecDist}. Final discussions and potential applications of the 
methods are presented in~Sec.\ref{SecConc}.

\section{Models and methods}
\label{SecModels}

\begin{figure}[!tb]
\includegraphics[width=7.5cm]{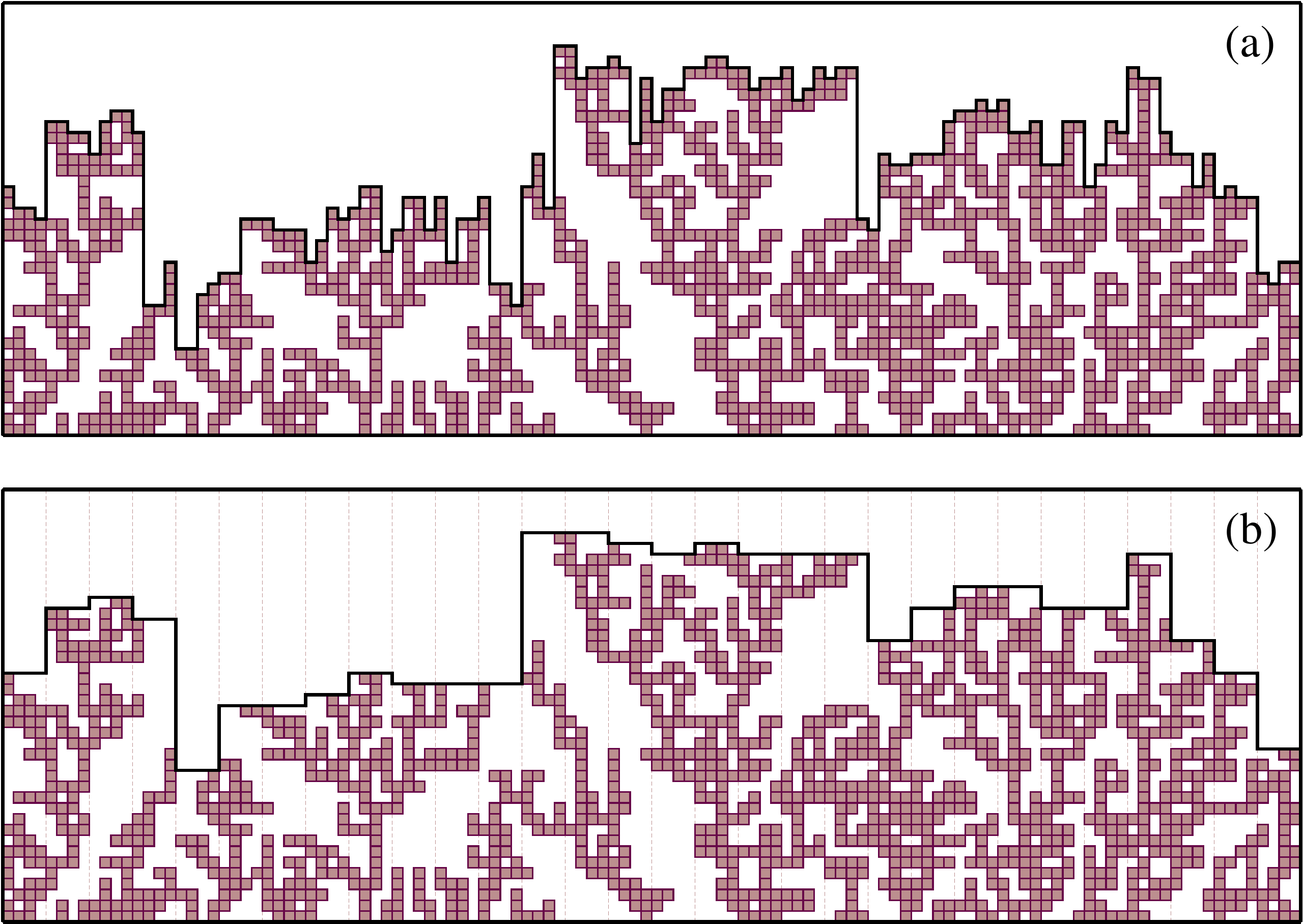}\\~\\
\includegraphics[width=7.5cm]{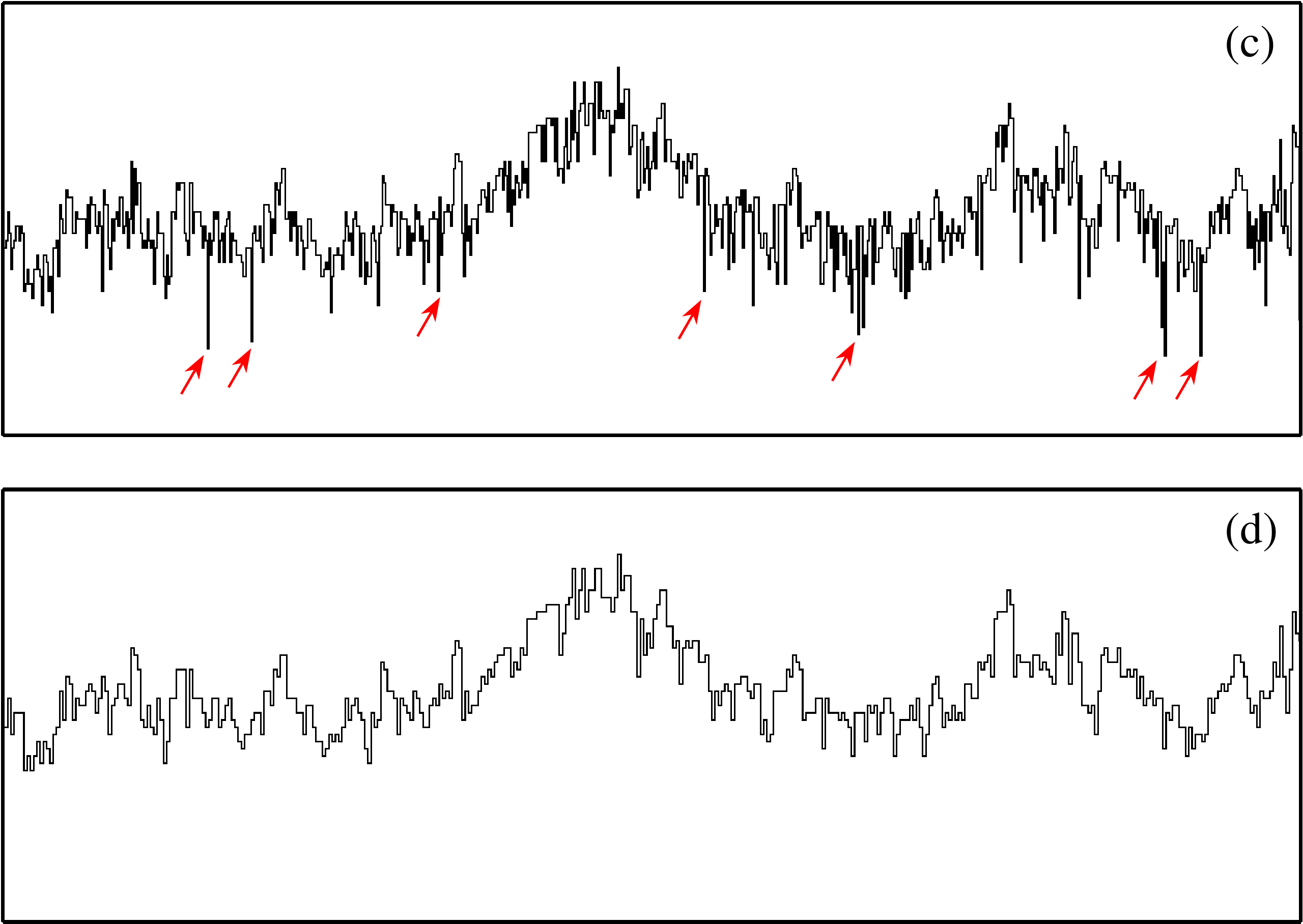}
\caption{(Color online) (a) A typical deposit of the BD model in $d=1+1$ and its 
height profile (for $\varepsilon=1$). (b) Height profile built using local 
maximal heights in boxes of size $\varepsilon=4$ for the same deposit of (a). 
The vertical dashed lines indicate the separation between boxes used to construct 
the coarse-grained profile. Typical cross-sections of standard and binned 
($\varepsilon=2$) surfaces for the BD model in $d=2+1$, with $L=800$ and $t=1000$,
are shown in panels (c) and (d), respectively. Arrows indicate a few narrow-deep 
valleys removed with the binning method. } 
\label{fig:models}
\end{figure}

In the ballistic deposition (BD) model~\cite{barabasi}, particles are randomly 
released perpendicularly to an initially flat substrate and permanently stick at 
their first contact with the deposit or the substrate. Therefore, porous 
deposits with large steps at the surface are formed, as shown in 
Fig.~\ref{fig:models}(a). 

We also investigated the grain deposition (GD) model~\cite{tiago1} conceived to 
simulate grained surfaces. In this model, cubic grains of side $l$ (in units of 
the lattice parameter) are released in a trajectory perpendicular and 
with two faces parallel to the substrate, at randomly chosen positions. The grains 
permanently aggregate when their bottoms touch the top of a previously deposited 
grain or the substrate. The deposited grain is usually laterally shifted in 
relation to underneath grains, which also leads to a porous deposit and large 
steps are formed in the surface~\cite{tiago1}. We present results for grain 
sizes $l=2$ and $l=4$, hereafter named as GD2 and GD4, respectively.

Both models are defined on square lattices with periodic boundary conditions. A 
unity of time is defined as the deposition of $L^2$ particles (lattice unitary cells) 
in both models. Therefore, in GD model, $L^2/l^3$ grains are deposited during a 
time unity. We study these models on square lattices of lateral sizes up to $L=2^{14}$. 

The surface of ballistic models is conventionally defined as the highest points 
of the deposit at each lattice position. With this standard definition the 
resulting surface have many narrow-deep valleys, as shown in 
Fig.~\ref{fig:models}(a) for BD model in $d=1+1$. These valleys are more 
pronounced in $d=2+1$, as shown in Fig.~\ref{fig:models}(c). In 
section~\ref{SecIntrinsic}, we will show that the fluctuations in the height 
increments during the deposition process are responsible by the strong 
corrections to the scaling in ballistic models. We propose that the leading 
contribution to these fluctuations is due to  these narrow-deep valleys.  In 
order to check this hypothesis, we introduce an alternative definition of the
surface considering only the largest local heights. More precisely, we divide 
the surface in boxes (bins) of size $\varepsilon$ and take only the maximal 
height inside each box to form a coarse-grained surface with $(L/\varepsilon)^2$ 
sites. Figures~\ref{fig:models}(b) and (d) show typical height profiles 
obtained with binned surfaces for BD model in $d=1+1$ and $2+1$, respectively. 
As expected, smoother surfaces are obtained since many narrow-deep valleys are 
discarded. The values of $\varepsilon$ must be small when compared with the 
typical size of the large wavelength fluctuations, which are responsible by the 
universality class of the system. Notice that, in fact, the binning procedure preserves 
the long wavelength fluctuations.

\section{Non-universal parameters}
\label{SecPara}

\begin{figure}[!t]
 \includegraphics[width=7.0cm]{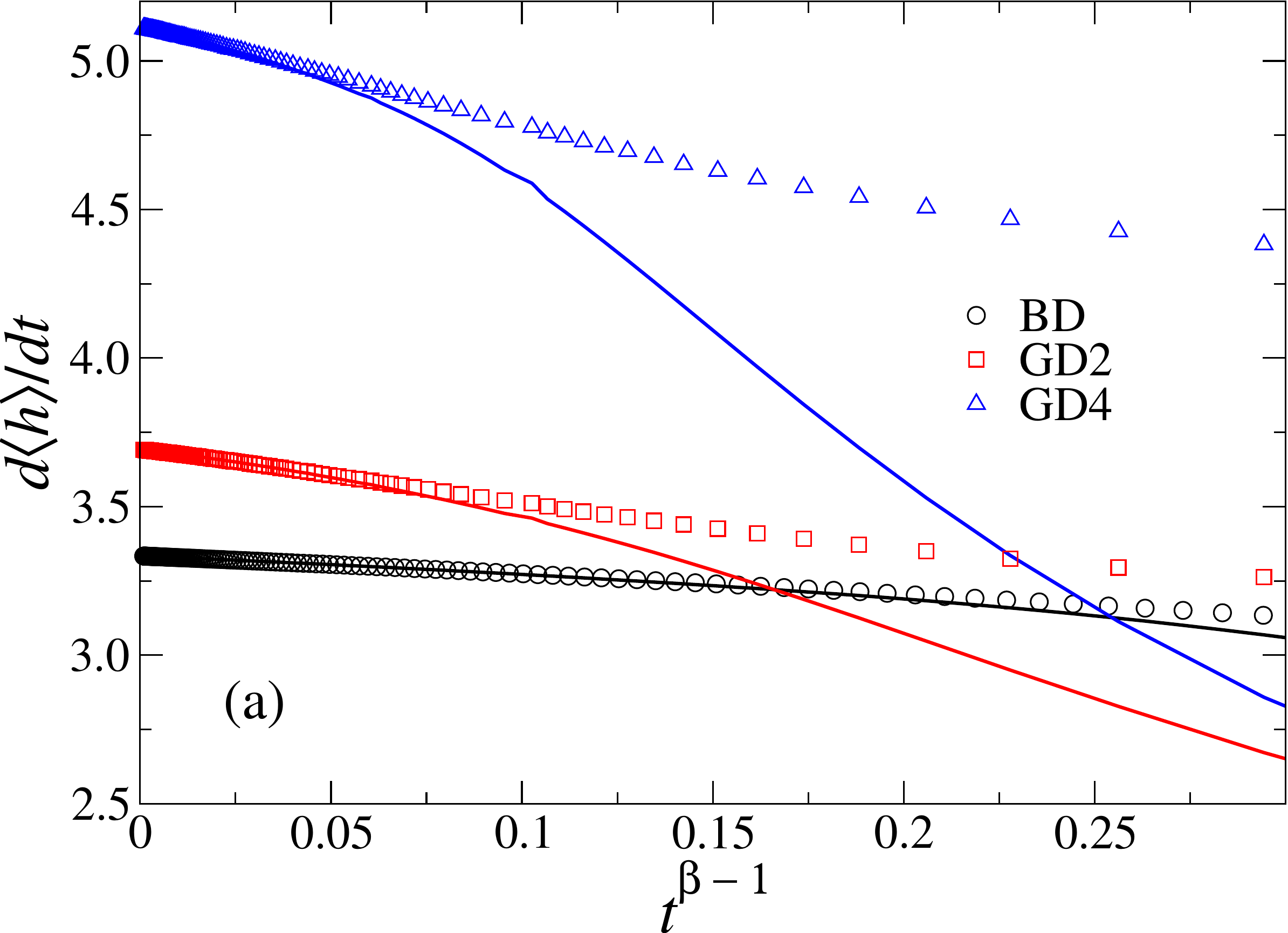} \\~\\
 \includegraphics[width=7.0cm]{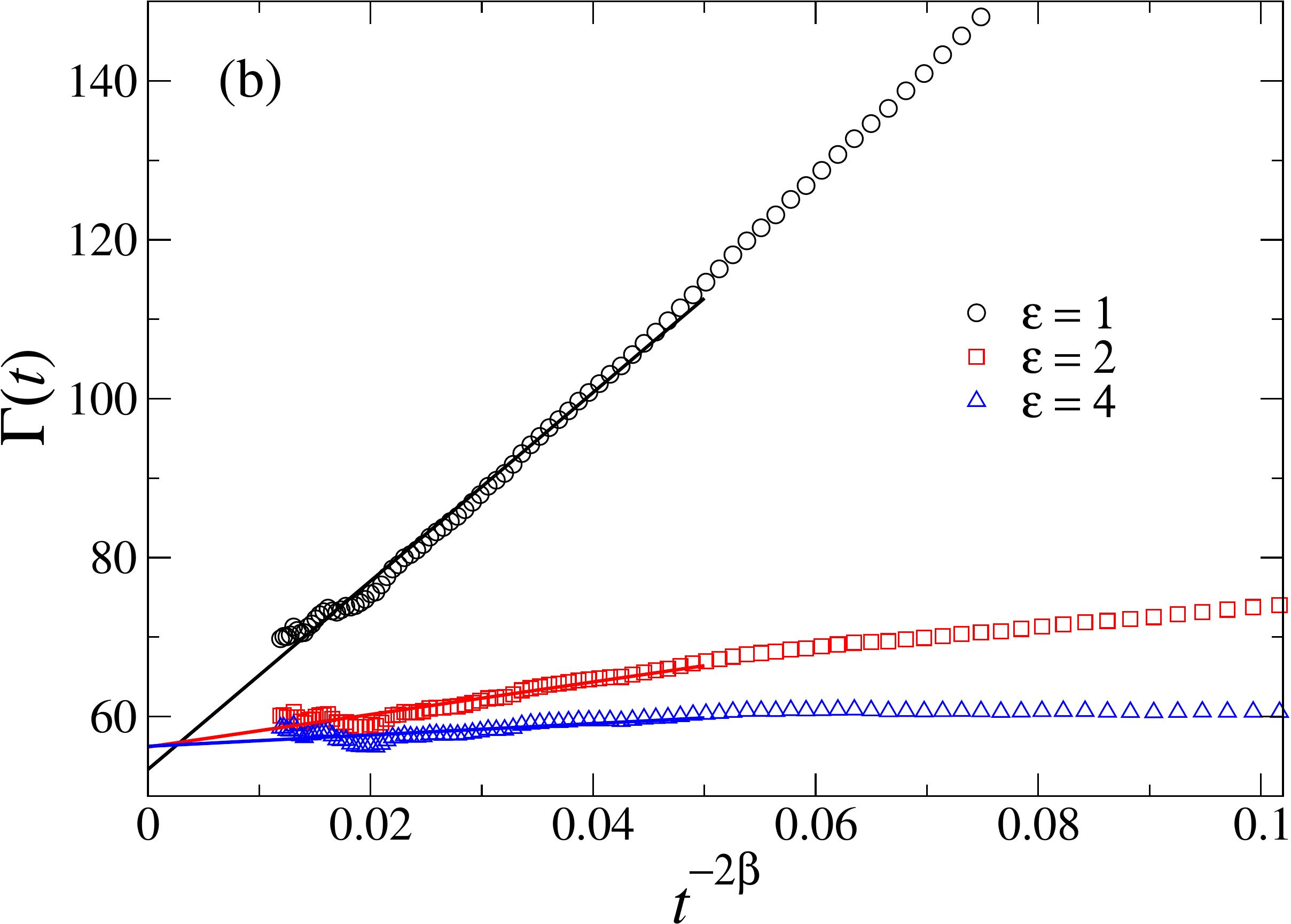}
\caption{(Color online) (a) Growth velocity against $t^{\beta-1}$ for distinct 
ballistic models. The surface was built using $\varepsilon$ twice the 
grain/particle size. Lines indicate the same quantities for the standard surface 
($\varepsilon=1$). (b) Amplitude fluctuation parameter estimated via 
KPZ ansatz for BD surfaces using distinct coarse-graining parameters. Lines are 
linear regressions to extrapolate $\Gamma$ in the limit $t\rightarrow\infty$.} 
\label{fig:vinf}
\end{figure}

The scaling analysis based on the KPZ ansatz, Eq.~(\ref{eqansatzcorr}), requires 
accurate estimates of the non-universal parameters. The growth velocity $v=d 
\lrangle{h}/d t$ against $t^{\beta-1}$ is shown for 
all studied models in Fig.~\ref{fig:vinf}(a), using both standard and binned surfaces definitions. Here, we 
adopt $\beta=0.241$ as the KPZ growth exponent in $d=2+1$~\cite{Kelling}. For 
the standard surface ($\varepsilon=1$), strong corrections are found and the 
linear regime expected in the KPZ ansatz is observed only for very long times. 
However, for $\varepsilon>1$ the linear regime is much more evident. The central 
point is that the convergence is faster for $\varepsilon>1$, but the asymptotic 
velocity does not depend on the value of $\varepsilon$.
The values of $v_{\infty}$ for the investigated models are shown in 
Table~\ref{tab:parameters}.

\begin{table}[!b]
\begin{tabular}{cccccccc}\hline\hline
Model  & & $v_\infty$  & & $\lambda$ & &$\Gamma$   \\\hline
BD     & & 3.33396(3)  & & 2.15(10)  & & 57(7)    \\
GD2    & & 3.6925(1)   & & 0.35(3)   & & 3.5(3)$\times 10^3$ \\
GD4    & & 5.1124(1)   & & 0.76(3)   & & 4.3(7)$\times 10^4$  \\ \hline\hline
\end{tabular}
\caption{\label{tab:parameters} Non-universal parameters for  ballistic models. }
\end{table}

The non-universal parameter controlling the amplitude of fluctuations in the KPZ 
ansatz can be obtained by the relation $\Gamma=|\lambda| 
A^{1/\alpha}$~\cite{krug90}, where $\alpha=0.393(4)$ is adopted as the roughness exponent 
for the KPZ class in $d=2+1$~\cite{Kelling}. The parameter $\lambda$ can be 
determined using deposition on tilted large substrates with an overall slope 
$s$, for which a simple dependence between velocity and slope, 
\begin{equation}
v \simeq v_\infty+\frac{\lambda}{2}s^2,
\label{eq:lambda}
\end{equation}
is expected for the KPZ equation~\cite{krug90}.
The parameter $A$ is obtained from the asymptotic velocity $v_L$ for
finite systems of size $L$~\cite{krug90} using the relation
\begin{equation}
 \Delta v = v_L-v_\infty \simeq -\frac{A\lambda}{2}L^{2\alpha-2}.
 \label{eq:vl}
\end{equation}
This approach is commonly called Krug-Meakin method~\cite{krug90} and the 
estimated values of $\lambda$ and $\Gamma$ parameters are shown in 
Table~\ref{tab:parameters}. Notice that, as mentioned before, asymptotic velocities are independent of 
the coarse-graining parameter $\varepsilon$ but the convergence is faster for 
larger $\varepsilon$. Thus, the parameters shown in Table~\ref{tab:parameters} 
were calculated for $\varepsilon=4 l$, where $l$ is the particle/grain size, which are the most accurate we obtained. Notice 
that we have $l=1$ for BD model, even though the GD model with $l=1$ is a random 
deposition \cite{barabasi}.

According to the extended KPZ ansatz, Eq.~(\ref{eqansatzcorr}), $\Gamma$ can also be obtained using
\begin{equation}
\Gamma=\lim_{t\rightarrow\infty} 
\left[\frac{\lrangle{h^2}_c}{t^{2\beta}\lrangle{\chi^2}_c}\right]^{1/2\beta},
\label{eqGamma}
\end{equation}
where $\lrangle{\chi^2}_c=0.235$ was adopted~\cite{healyPRL,healyPRE}.
In the case of $\eta$ independent of $\chi$, it is easy to check that 
Eq.~(\ref{eqansatzcorr}) implies
\begin{equation}
\Gamma(t) \equiv \left[\frac{\lrangle{h^2}_c}{t^{2\beta}\lrangle{\chi^2}_c}\right]^{1/2\beta} 
= \Gamma(\infty)+ct^{-2\beta}+\cdots.
\end{equation}
Fig.~\ref{fig:vinf}(b) confirms that $\Gamma(\infty)$ is independent of 
$\varepsilon$ and also that the correction in $\Gamma(t)$ is consistent with 
$t^{-2\beta}$. The asymptotic $\Gamma$ values obtained using this approach are 
the same, inside errors, as those found using the Krug-Meakin analysis shown in 
Table~\ref{tab:parameters}. In summary, we conclude that the role played by 
corrections in Eq.~(\ref{eqansatzcorr}) is being suppressed in surfaces for 
$\varepsilon>1$. 

\section{Scaling corrections and the intrinsic width}
\label{SecIntrinsic}

\begin{figure}[!t]
 \centering
 \includegraphics*[width=7cm]{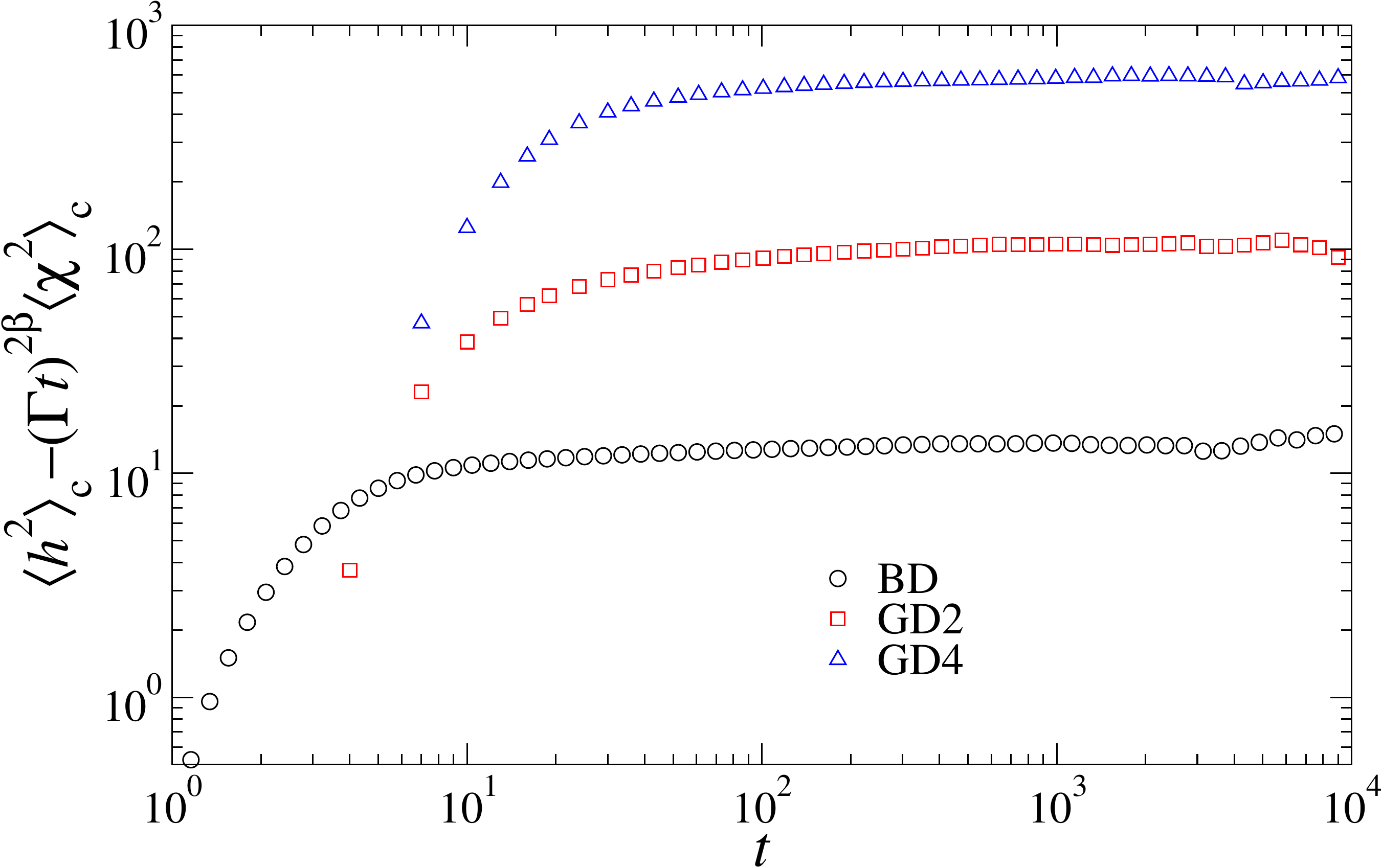} ~
 \caption{Determination of the intrinsic width for ballistic models 
using the KPZ ansatz. These curves were obtained using 
$\lrangle{\chi^2}_c=0.235$ and $\Gamma=57$, 3500 and 43000 for BD, GD2 and GD4 models, 
respectively.
}
 \label{fig:wint}
\end{figure}

The finite-time corrections in Eq.~(\ref{eqansatz}) are non-universal and 
its nature, deterministic or stochastic for example, will depend on the 
investigated model~\cite{Frings,Alves13}. The second cumulant 
is, for the most general case, given by
\begin{equation}
\lrangle{h^2}_c = (\Gamma t)^{2 \beta} \lrangle{\chi^2}_c+ 2(\Gamma t)^\beta 
\mbox{cov}(\chi,\eta) + \lrangle{\eta^2}_c + \cdots
\label{eqroug2}
\end{equation}
where $\mbox{cov}(\chi,\eta)\equiv\lrangle{\chi\eta}-\lrangle{\chi}\lrangle{\eta}$ is the covariance. 
In order to determine the relevant corrections in $\lrangle{h^2}_c$, we plot $\lrangle{h^2}_c-(\Gamma t)^{2 \beta} \lrangle{\chi^2}_c$
against time, as shown in Fig.~\ref{fig:wint}. The corrections reach a 
constant value - the squared intrinsic width $w_i^2$ - at relatively short 
times, ruling out a statistical dependence between $\chi$ and $\eta$, i.e., 
$\mbox{cov}(\chi,\eta)=0$. At first glance,  one would  identify 
$w_i^2=\lrangle{\eta^2}_c$ from  Eq.~(\ref{eqroug2})  but, in principle, the 
contribution of higher order corrections to $w_i^2$ cannot be disregarded.

We define the squared intrinsic width as 
\begin{equation}
w_i^2=\lim_{1\ll t \ll L^z}\left[\lrangle{h^2}_c-(\Gamma t)^{2 \beta} \lrangle{\chi^2}_c\right]. 
\label{eq:wiKPZ}
\end{equation}
The values  $w_i = 3.6(2)$, 10.0(5), 26(2) were obtained for BD,  GD2 and GD4 models, 
respectively. These estimates are in good agreement with the intrinsic 
width determined for these models using the collapse of the interface width 
distributions in the steady state ($t\gg L^z$)~\cite{tiago2}, implying that the 
intrinsic width formed in the dynamic regime ($t\ll L^z$) lasts indefinitely.

\begin{figure}[!b]
 \centering
 \includegraphics[width=7cm]{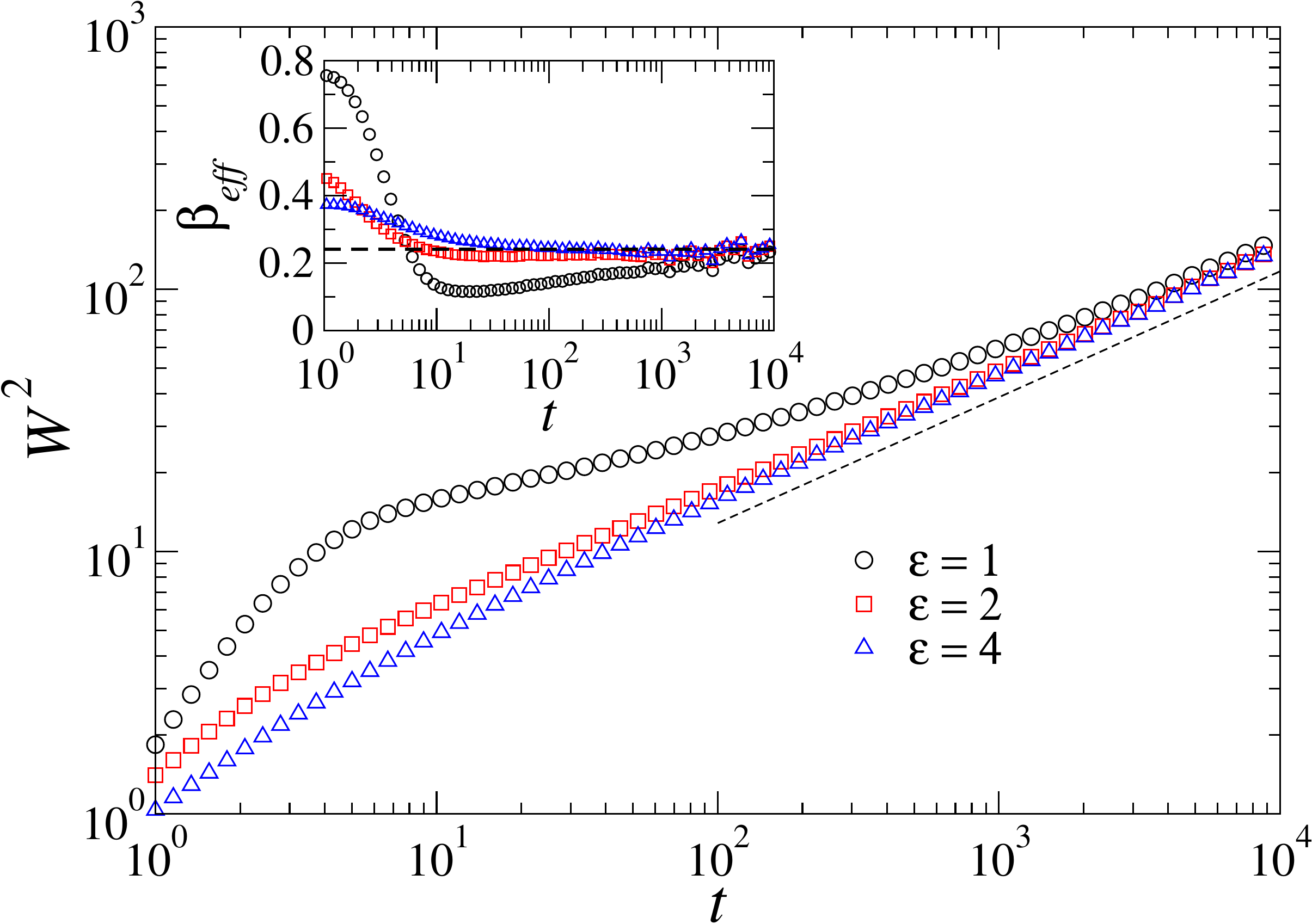}
 \caption{Squared interface width for BD surfaces using different  
binning parameters. Dashed line is a power law with exponent 
$2\beta_{kpz}$. Effective growth exponents are shown in the inset. The horizontal 
line represents $\beta_{kpz}=0.241$~\cite{Kelling}.
}
 \label{fig:w2dbeps}
\end{figure}

The squared interface width against time obtained for BD model using different 
binning parameters are shown in Fig.~\ref{fig:w2dbeps}. A quick convergence  to 
the KPZ scaling is found when $\varepsilon>1$ is considered and a large 
intrinsic width seems to be absent. This result is corroborated by the effective 
growth exponents $\beta_{eff}$, defined as the local slopes in 
double-logarithmic plots of $W$ against $t$, shown in the inset of 
Fig.~\ref{fig:w2dbeps}. Indeed, the intrinsic width determined in plots 
equivalent to Fig.~\ref{fig:wint}(a) results in a reduction from $w_i=3.6(2)$ 
for standard surface to $w_i=1.5(2)$ and $0.8(2)$ when $\varepsilon=2$ and 4 are 
used, respectively. Similar results were found for the GD models.

\begin{figure}[!t]
 \centering
 \includegraphics[width=7cm]{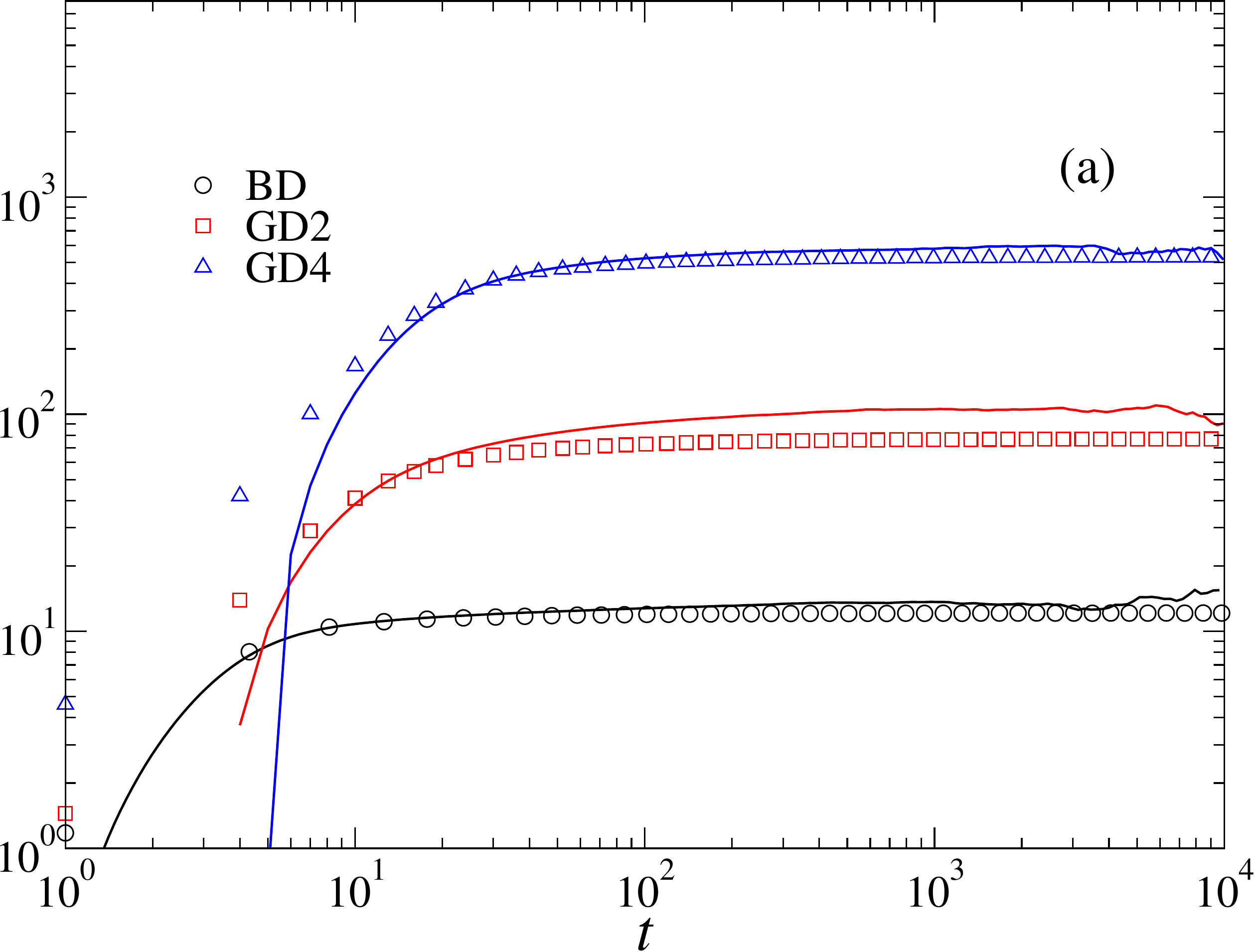} ~
 \includegraphics[width=7cm]{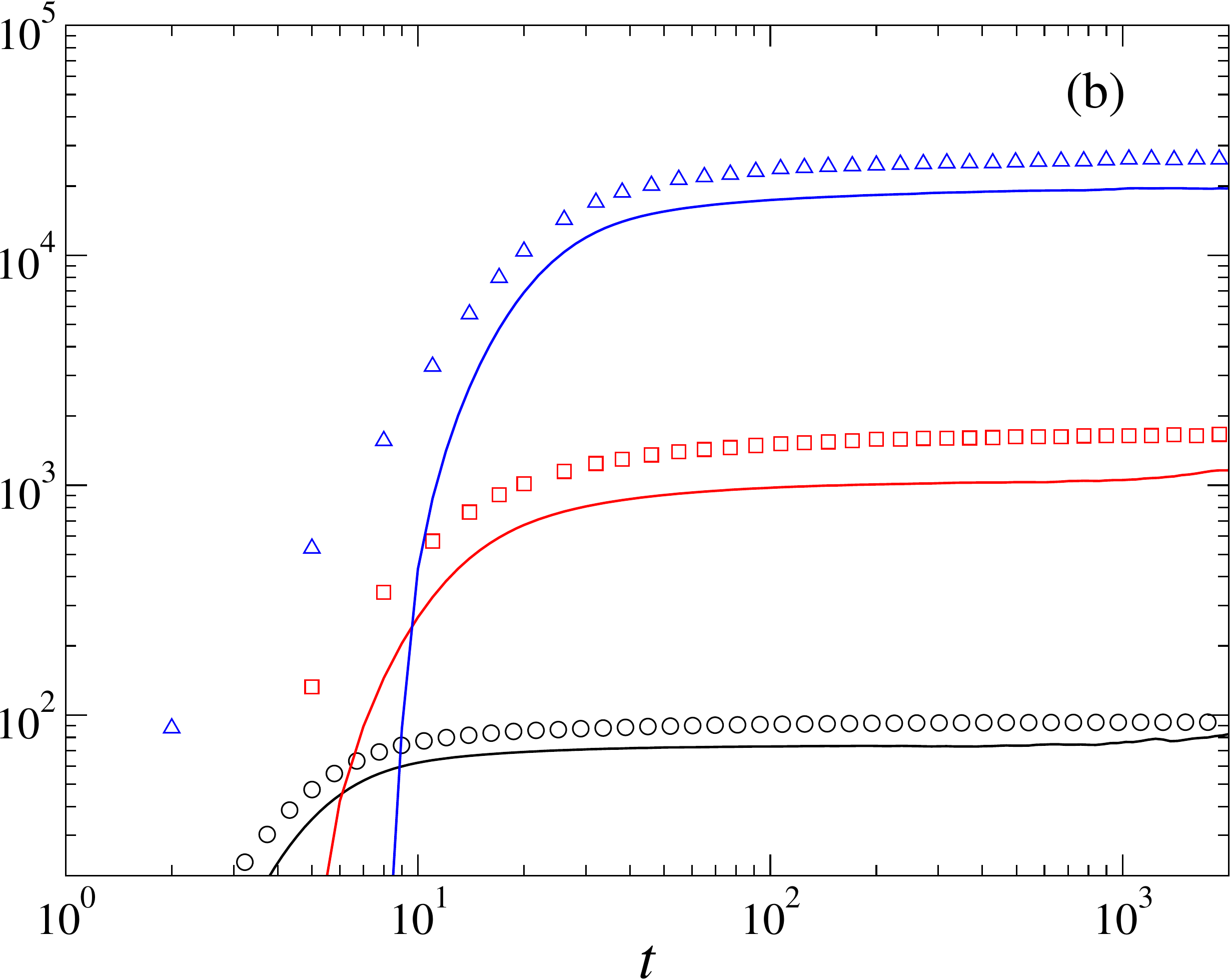}
 \caption{(a) Second order cumulant of $\delta h$ (symbols) and 
$\lrangle{h^2}_c-(\Gamma t)^{2 \beta} \lrangle{\chi^2}_c$ (lines) against time. 
(b) Third order cumulant of $\delta h$ (symbols) and $\lrangle{h^3}_c-(\Gamma t)^{3 
\beta} \lrangle{\chi^3}_c$ (lines) against time.} 
\label{fig:dh}
\end{figure}

We conclude that the presence of narrow-deep valleys is, in fact, a necessary 
condition to observe strong corrections and, consequently, the intrinsic width. In 
these valleys with large steps, the heights are incremented, in average, by 
$\lrangle{\delta h}\rightarrow l v_\infty>1$, where the factor $l$ (particle/grain 
size) is required to account for the time step definition in GD models. However, these 
increments are not deterministic and we propose that their stochastic 
fluctuations are the leading contributions to the intrinsic width. In order to 
validate this conjecture, we determined $\lrangle{(\delta h)^2}_c$, where 
$\delta h=h(\mathbf{x},t+\delta t)-h(\mathbf{x},t)$ is the height increment in a 
time step. The results are compared with the squared intrinsic width, obtained 
via the KPZ ansatz, in Fig.~\ref{fig:dh}(a). For all models, the $w_i^2$ is slightly larger than 
$\lrangle{(\delta h)^2}_c$. Since the binning method reduces the intrinsic width,
we conclude that the leading contribution to $w_i^2$ comes
from the fluctuations in narrow-deep valleys.

A central contribution to the mechanism behind the leading corrections in  
ballistic growth models is therefore elicited. However, it is not exclusivity of 
ballistic models. For example, an intrinsic width can also be  
determined for the RSOS model~\cite{KK}. In this model, at each time step, the 
height of a randomly selected column is incremented by a unity if the height 
difference between nearest neighbors obeys the constraint $|h_j-h_{j'}|\le m$, 
otherwise, the deposition attempt is refused. This model produces an asymptotic 
growth velocity $v_\infty<1$ independently of the substrate 
dimension~\cite{Oliveira12,Oliveira13R,Alves13}. Since only increments $\delta 
h=1$ or 0 are allowed, we have that deposition and refusal for long times occur 
with probabilities $v_\infty$ and $1-v_\infty$, respectively. Therefore, we have 
$\lrangle{(\delta h)^2}_c=v_{\infty}(1-v_{\infty})$. In $d=1+1$ and 
$2+1$, the RSOS asymptotic velocities for $m=1$ are 
$v_{\infty}\approx0.419$~\cite{Oliveira12} and 0.3127~\cite{Oliveira13R} 
resulting in small variances $\lrangle{(\delta h)^2}_c\approx 0.24$ and 0.21, 
respectively.  We simulated the RSOS model and found $w_i^2\approx 0.20(15)$ 
using Eq.~\ref{eq:wiKPZ} for both dimensions. The RSOS model also helps to 
understand why the intrinsic width cannot be solely associated to large steps in 
surface. One can chose a large value of $m$ such that steps of the same order of 
the ones in the BD are present in RSOS interfaces. However, one  still has  
$\lrangle{(\delta h)^2}_c < 1$ irrespective of $m$, which introduces a small 
correction in the  scaling. 

Figure \ref{fig:dh}(b) shows the corrections in the third cumulant of heights, 
$\lrangle{h^3}_c-(\Gamma t)^{3 \beta} \lrangle{\chi^3}_c$, against time, where 
$\lrangle{\chi^3}_c=0.049$ was estimated using data given in 
Refs.~\cite{healyPRL,healyPRE,Oliveira13R}. Again, the main correction is a 
constant that is smaller than $\lrangle{(\delta h)^3}_c$. 
The same behavior was found in the fourth order cumulants. So, based on these data 
we clearly show a strong correlation between finite-time corrections in 
the KPZ ansatz and $\delta h$ but we could not infer a simple functional 
dependence.

\begin{figure}[!t]
 \centering
 
 \includegraphics[width=7cm]{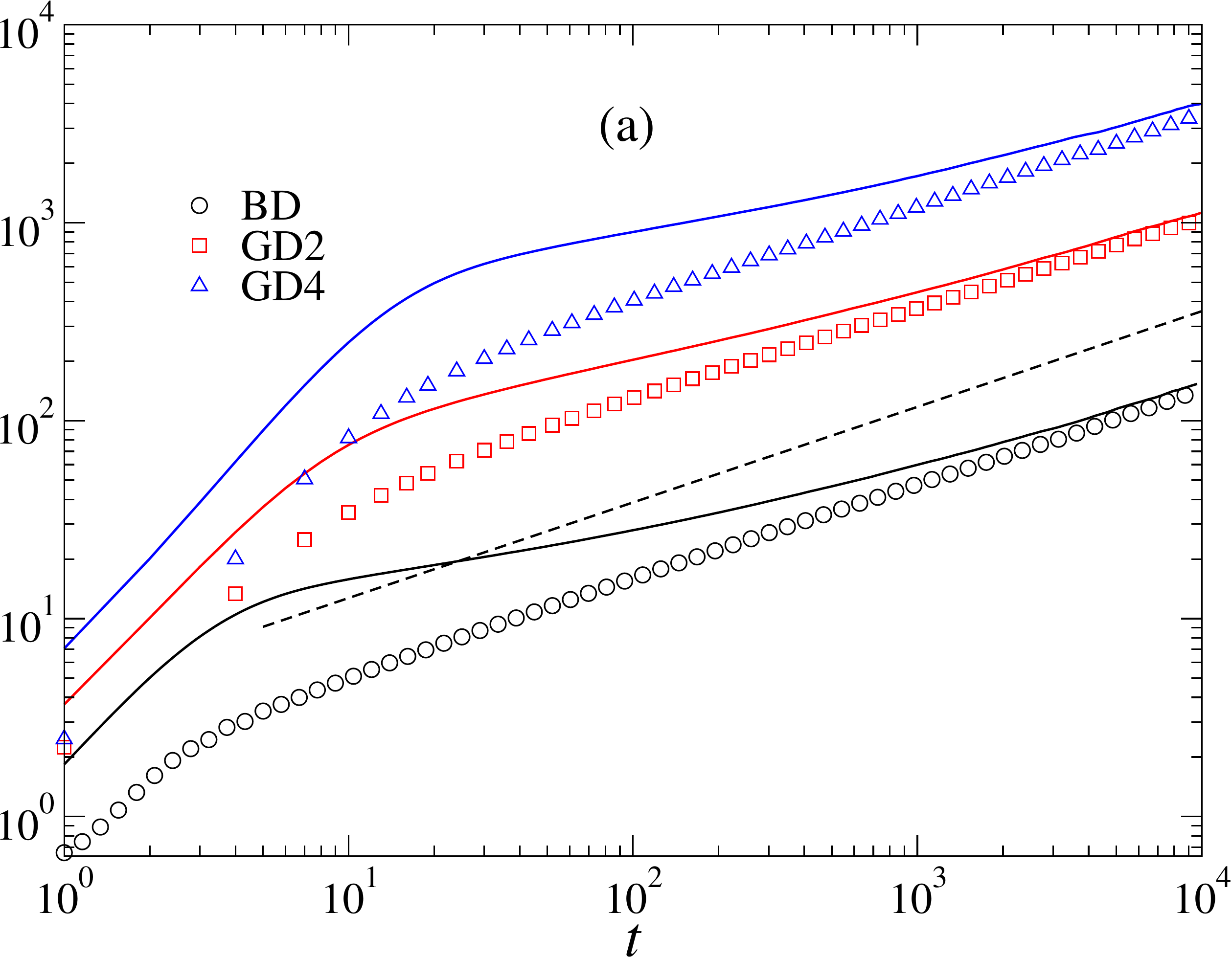}
 
 \includegraphics[width=7cm]{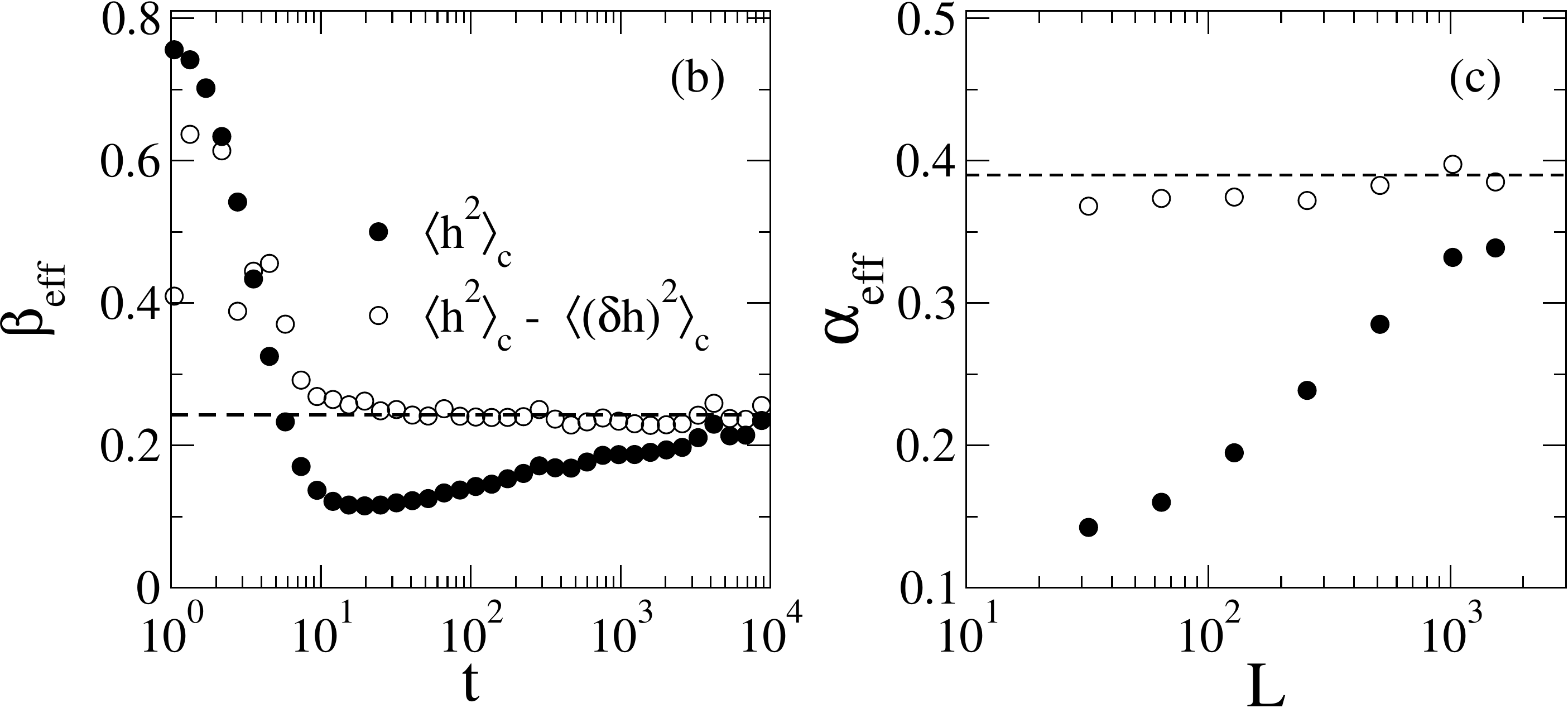}
 \caption{(a) Squared interface width subtracted (symbols) or not (continuous lines) 
of $\lrangle{(\delta h)^2}_c$  against time for distinct ballistic models. 
The dashed line is a power law with exponent 
$2\beta_{kpz}=0.483$~\cite{Kelling}. (b) Effective growth exponent against time 
for BD model considering (open symbols) or not (filled symbols) the intrinsic 
width. (c) Effective roughness exponent against size for BD model.}
 \label{fig:eff}
\end{figure}

The interface width against time for $\varepsilon=1$, discounting or not the 
$\lrangle{(\delta h)^2}_c$, is shown in Fig.~\ref{fig:eff}(a). While the 
original curves do not scale as a power law for the investigated times (solid 
lines), the subtraction of $\lrangle{(\delta h)^2}_c$  leads to an excellent 
accordance with the growth exponent of the KPZ class, even for relatively short 
times. This analysis is confirmed through the effective growth exponents, 
obtained from either $W^2$ vs. $t$ or $W^2-\lrangle{(\delta h)^2}_c$ vs. $t$, 
shown in Fig.~\ref{fig:eff}(b). Similar plots are found for GD models. The 
obtained exponents are shown in Table~\ref{tab:uni} and are in remarkable 
agreement with the best estimates of the KPZ growth exponent $\beta=0.2415(10)$ 
in d=2+1~\cite{Kelling}. The exponents are, inside errors, the same as those 
obtained for binned surfaces built with $\varepsilon>1$. 

The effective roughness exponents for ballistic deposition including or not the 
$\lrangle{(\delta h)^2}_c$ are shown in Fig.~\ref{fig:eff}(c) and the estimates 
are given in Table~\ref{tab:uni}. The results for BD are 
again in very good agreement while those for GD models are slightly below the best 
estimates  for KPZ exponents in $d=2+1$, $\alpha=0.393(3)$~\cite{Kelling},  in 
sharp contrast with a very poor accordance obtained when the intrinsic width is 
disregarded. Indeed, if we neglect intrinsic width and use only $1024\le L\le 
2048$, the exponents are $\alpha\approx0.33$, 0.32 and 0.26 for BD, GD2 and GD4, 
respectively. The exponents constitute a strong evidence that these models 
belongs, in fact, to the KPZ universality class in $d=2+1$.

\section{Height distribution analysis}
\label{SecDist}

\begin{figure}[!t]
 \centering
 \includegraphics[width=8cm]{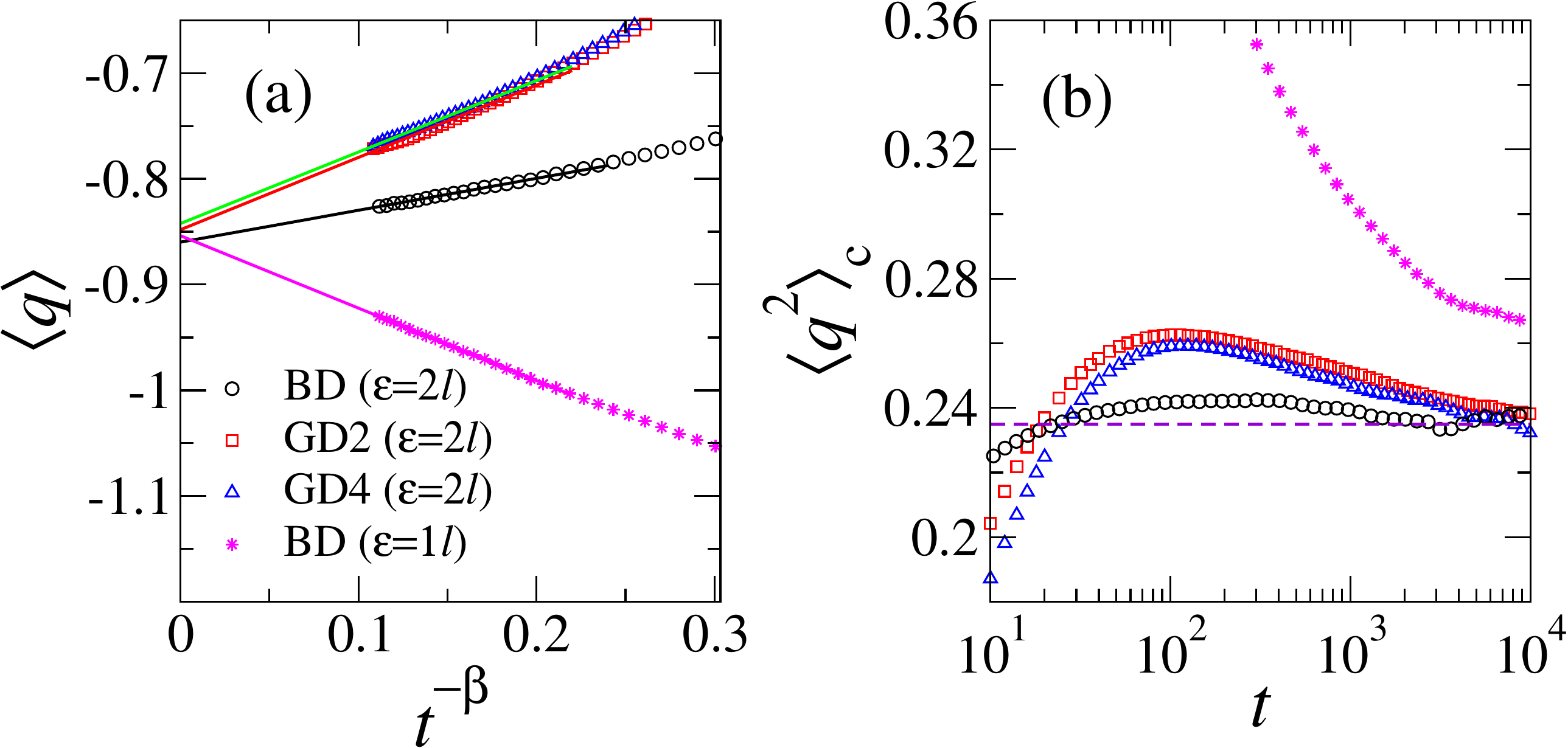}\\
 ~\\
 \includegraphics[width=7cm]{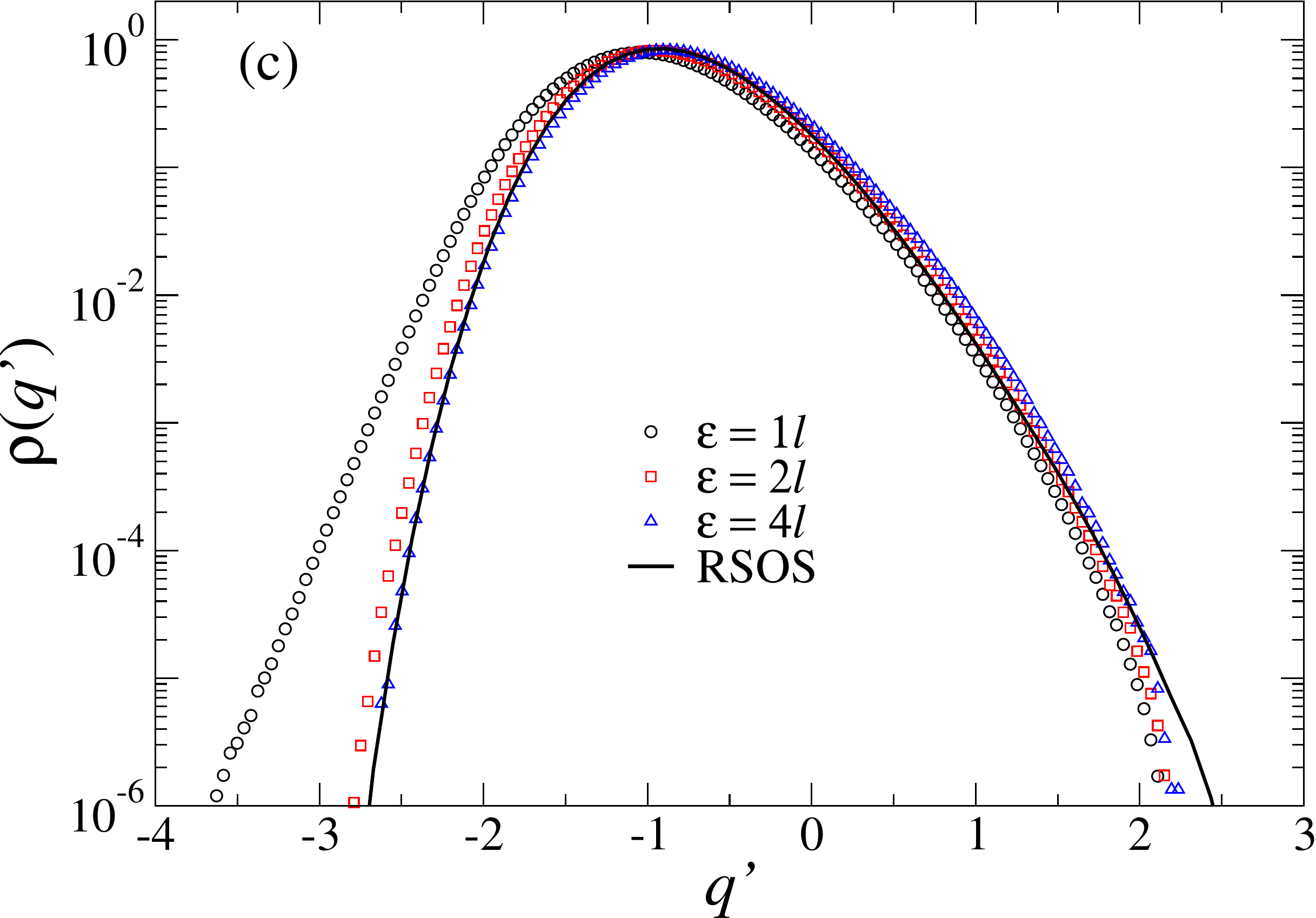}
 \caption{Determination of the (a) first and (b) second cumulants of the stochastic quantity
 $\chi$. Solid lines are linear regressions to extrapolate $\lrangle{\chi}$
 and the dashed one represents $\lrangle{\chi^2}_c=0.235$.
 (c) Rescaled height distributions for BD model at deposition time $t=10^4$. 
 Solid line is the distribution obtained for the RSOS model~\cite{Oliveira13R}. 
 Here $q'=(h-\vinf t-\lrangle{\eta})/(\Gamma t)^\beta$.}
 \label{fig:rhochi_db}
\end{figure}

The presence of narrow-deep valleys at surface of ballistic 
deposition model is a hindrance to check the 
universality of the stochastic quantity $\chi$ in the KPZ ansatz. Thus, height 
distributions were analyzed using surfaces built with $\varepsilon>1$. The first 
and second cumulants of $\chi$ can be obtained analyzing the asymptotic value of 
$\lrangle{q}$ and $\lrangle{q^2}_c$, where $q$ is defined by Eq.~(\ref{eq:q}). 
The results obtained for $\varepsilon=2l$ are shown in 
Figs.~\ref{fig:rhochi_db}(a) and \ref{fig:rhochi_db}(b). The results are 
essentially the same for $\varepsilon=4l$. Since the standard 
correction $t^{-\beta}$ is present in the first cumulant, a extrapolation in 
time using the proper power law is imperative for a reliable 
estimate in finite time simulations~\cite{Oliveira13R}. The  extrapolated values are shown in 
Table~\ref{tab:uni} and agree, inside errors, with the best estimates known for 
the KPZ class in $d=2+1$~\cite{healyPRE,alves14}. For sake of comparison, results 
for BD model with $\varepsilon=1$ are also shown in Figs.~\ref{fig:rhochi_db}(a) 
and \ref{fig:rhochi_db}(b). In the former, we observe that the asymptotic 
estimate of $\lrangle{\chi}$ is almost independent of $\varepsilon$, but the 
mean value of the correction $\lrangle{\eta}$ is strongly affected by the choice of 
$\varepsilon$, resulting the values $\lrangle{\eta}=-1.70$, 0.94 and 2.60 for 
$\varepsilon=1$, 2 and 4, respectively. In the latter, the curve for 
$\varepsilon=1$ apparently converges towards the KPZ value, but still far 
from it even at the longest analyzed time.

The dimensionless cumulant ratios skewness $S=\lrangle{h^3}_c/\lrangle{h^2}_c$ 
and kurtosis $K = \lrangle{h^4}_c/\lrangle{h^2}_c^2$, calculated in the growth 
regime, are universal and in agreement  with the known values for the KPZ class 
in $d=2+1~$\cite{healyPRL,healyPRE,Oliveira13R}, as one can see in  
Table~\ref{tab:uni}. We also analyzed the skewness of the height distributions 
in the steady state. For $L \ge 1024$ and $\varepsilon \ge 2l$, we found $S$ in 
agreement, within the uncertainties, with the value $S=0.26(1)$ estimated from 
other KPZ models that have small corrections to the 
scaling~\cite{fabioBD,Marinari,chin}. Notice that negative skewed stationary distributions 
were reported in Ref.~\cite{tiago2} using the standard surface definition 
($\varepsilon=1$) and the same sizes considered here. However, $S>0$ is expected 
in ballistic models since $\lambda> 0$, which was indeed found in our analysis 
with $\varepsilon>1$. Our results show that our method is also able to strongly 
reduce the finite-size corrections to scaling in the steady state. 

\begin{table}[!t]
 \begin{tabular}{ccccccc}
  \hline\hline 
  model &  $\beta$  &  $\alpha$&  $\lrangle{\chi}$ &  $\lrangle{\chi^2}_c$ & S         &  K \\ \hline 
  BD    &  0.239(15)&  0.389(3)& 0.86(2)          &  0.235(15)            &  0.41(2)   &  0.31(3) \\
  GD2   &  0.225(15)&  0.375(5) & 0.85(2)         &  0.24(2)              &  0.43(3)   &  0.32(3) \\
  GD4   &  0.237(18)&  0.375(15) & 0.84(3)        &  0.24(2)              &  0.44(3)   &  0.35(5) \\
  \hline\hline
 \end{tabular}
\caption{\label{tab:uni} Universal quantities determined for ballistic growth models 
either discounting the intrinsic width ($\beta$ and $\alpha$) or using surfaces 
constructed with $\varepsilon=2l$ (other quantities). Uncertainties in 
cumulants and cumulant ratios were obtained propagating the uncertainties in the 
non-universal parameters $v_\infty$ and $\Gamma$ given in Table 
\ref{tab:parameters}.}
\end{table}

The height distributions rescaled according to the KPZ ansatz are shown in Fig. 
\ref{fig:rhochi_db}(c) for the BD model after a growth time $t=10^4$ using different 
binning parameters. This figure also shows the distribution obtained for 
the  RSOS model that has small corrections to the scaling and exhibits excellent 
agreement with the KPZ ansatz in $d=2+1$ 
dimensions~\cite{healyPRL,healyPRE,Oliveira13R}. The distribution for the 
standard surface exhibits strong deviations in the left tail associated to 
fluctuations below the mean height (since $\lambda>0$), where  deep valleys 
contributions are present. The rescaled distributions for binned surfaces  are 
very close to the RSOS one.  Therefore, we show that ballistic growth models 
in $d=2+1$ dimensions obey the  KPZ ansatz with the expected universal 
stochastic term $\chi$, which would be practically impossible with the currently 
computer resources if the strong finite-time corrections were not explicitly
taken into account in the analysis.

\section{Final discussions and conclusions}
\label{SecConc}

In summary, we have showed that the leading corrections to the scaling 
of ballistic growth models in $d=2+1$ arise from the large stochastic 
fluctuations of the height increments  $\delta h$ during the deposition process, 
which is expressed in the form of an intrinsic width $w_i$ in the Family-Vicsek 
scaling, Eq.~(\ref{eqFVg}). We observed that $w_i^2 \approx \lrangle{(\delta 
h)^2}_c$. This intrinsic width also exists in solid-on-solid KPZ models, but in 
this case $w_i^2 \approx \lrangle{(\delta h)^2}_c < 1$, so that corrections to 
scaling are negligible. Anyway, since the variance $\lrangle{(\delta h)^2}_c$ 
can be easily computed in numerical simulations, we propose that it should be 
calculated together with the squared interface width $W^2$ and the standard 
analysis of $W^2$ against time $t$ or substrate size $L$, used in hundreds of 
previous works, should be replaced by  $W^2(t) - \lrangle{(\delta h)^2}_c$ 
\textit{vs}. $t$ or $L$, respectively. This procedure is able to eliminate the intrinsic 
width from the scaling analysis and to access the universal scaling exponents 
with affordable computer resources. We believe that this recipe must be a
standard in numerical studies of growing interfaces, helping to uncover the 
universality of  models where strong corrections play an important role.

The large fluctuations in the height increments $\delta h$ 
arises mainly from the aggregation at narrow-deep valleys in the surface. Therefore, we 
also propose a simple method that eliminates these valleys at the surface, 
where the leading contributions to these fluctuations take place. Basically, the 
original surface is binned in boxes of lateral size $\varepsilon$ and only the 
highest point inside each box is used to construct a coarse-grained surface and to perform 
statistics. We showed that, for $\varepsilon$ larger than the typical 
particle/grain size, the intrinsic width is strongly reduced.

Both methods yield scaling exponents for ballistic growth models in excellent 
agreement with the KPZ class in $d=2+1$. Despite of ballistic growth models 
present the requisites for KPZ class, to our knowledge, we provide the first 
convincing observation of KPZ scaling exponents in these models in two-dimensions. 
Moreover, the power of the binning method is not restricted to scaling exponents. Indeed, the 
effects of finite-time corrections in the KPZ ansatz become negligible if surfaces are built 
with $\varepsilon>l$, while the fundamental non-universal parameters [growth 
velocity $v_{\infty}$ and amplitude of fluctuations $\Gamma$ in 
Eq.~(\ref{eqansatz})] as well the universal quantities [growth exponent $\beta$ 
and $\chi$ in Eq.~(\ref{eqansatz})] remain unchanged. So, we showed that the 
rescaled height distributions for the growth regime of ballistic growth models 
are the same as those obtained for other KPZ models in $d=2+1$. Furthermore, the 
skewness of height distributions in the  steady state also shows a good 
agreement with the value accepted for the KPZ class in this dimension. Therefore, we 
show that the ballistic growth models in $d=2+1$ belongs to the KPZ 
universality class. In particular, our result ends a longstanding discussion about the 
validity of the KPZ class in the classic BD model~\cite{raissa,fabioBD,vvdensky}.

The binning  method can, in principle,  be easily applied in the analysis of 
experimental surfaces. As an example, consider the recent experiment by Yunker 
\textit{et al}. \cite{yunker}, where particles from a colloidal suspension were deposited 
at the edges of evaporating drops. For small anisotropy of the particles, the 
system was observed to be in the KPZ class. On the other hand, for highly 
anisotropic particles, exponents different from the KPZ class were found and 
attributed to the quenched KPZ class. However, this conclusion 
have been questioned~\cite{nicoli,tiago3}. In Ref.~\cite{nicoli} a transient 
anomalous scaling was proposed as a possible explanation for the deviation from the 
KPZ regime while in Ref.~\cite{tiago3} an advection-diffusion model
with strong corrections to the scaling due to a large intrinsic width was used 
to explain the deviation. In particular, we applied the binning method to 
advection-diffusion model of Ref. ~\cite{tiago3} and observed excellent
agreement with KPZ 
exponents (data not shown). Depending on the parameters, both model and 
experimental surfaces present a large number of narrow-deep valleys. Thus, we 
believe that our binning method can be very useful to solve controversial issues as  
the colloidal deposition problem and others related systems.

\begin{acknowledgments}

The authors acknowledge the support from CNPq and FAPEMIG (Brazilian agencies).

\end{acknowledgments}

\end{document}